D. Santamargarita et al., "Study of Applicability of Simple Closed Loop Input Impedance Model for Grid-Tie Inverters," IECON 2019 - 45th Annual Conference of the IEEE Industrial Electronics Society, Lisbon, Portugal, 2019, pp. 2026-2031.

DOI: 10.1109/IECON.2019.8926942

# Study of Applicability of Simple Closed Loop Input Impedance Model for Grid-Tie Inverters

D. Santamargarita[1], F. Huerta[1], M. Sanz[1], A. Lazaro[1], S. D'Arco[2], S. Sanchez[2], E. Tedeschi[3], J. Roldán[4]

[1] Universidad Carlos III de Madrid, Grupo de Sistemas Electrónicos de Potencia – Leganés, Madrid, Spain
[2] SINTEF – Trondheim, Norway
[3] Norwegian University of Science and Technology, NTNU – Trondheim, Norway
[4] Instituto IMDEA Energía, Unidad de Sistemas Eléctricos – Móstoles, Madrid, Spain
Email: dsantama@ing.uc3m.es

*Abstract*— In recent years the need for DC distribution buses has increased considerably. As it can be noticed in the transport for example the distribution systems of the more electric aircrafts, ships, or electric cars. Given the complexities of the systems presented above, the need to use more and more switched power converters has arisen. The main problem of the connection of multiple controlled switched converters acting as source and load is the degradation of stability that occurs on the DC distribution bus due to the converter interactions. To study the stability in the distribution bus there are some well-established criteria. These criteria require knowledge of the input impedance of the converters that act as load and the output impedance of the equipment that acts as source. In order to reduce the complexity of obtaining the input impedance a model based on a controlled converter acting as a constant power load (CPL) is commonly used. This article studies the accuracy of this model for a commonly used topology in distribution systems nowadays, Two Level Voltage Source Converter (2L-VSC), studying different scenarios that make the model become inaccurate.

*Keywords*— *Stability, Impedance, Reduced Order Model.*

## I. Introduction

In recent years, traditional energy distribution systems, which are based on a centralized architecture, have been progressively replaced by more distributed power systems, in which the number of interconnected power converters is increased. Some examples of these distribution systems can be seen in transport as the powertrain distribution systems of electric cars [1], power distribution for more electric aircraft [2] or power distribution for all-electric ships [3]. These distribution systems are also used in energy distribution systems with greater power, such as HVDC offshore wind farms [4].

One of the main problems of connecting different regulated power converters to the same DC distribution bus is the degradation of stability that can occur due to the interactions among the converters. Although each subsystem is independently designed to be standalone stable. There are several methods for determining systems interconnection stability, including the Middlebrook criterion or the Gain Margin and Phase Margin criterion (GMPM) [5]. All of the above-mentioned criteria require the use of small signal input and output impedance models since the connection of two individually stable systems can be modeled as the serial connection of the small signal output impedance of the source system $Z_S$ and the small signal input impedance of the load system $Z_L$ Fig. 1 the stability in this interconnection is determined by the open loop expression (1). In order to do the stability analysis in this paper the GMPM criterion will be used. The GMPM criterion ensures that when the conditions of (2) are met the connection of the two systems will be stable, these conditions are equivalent to the Nyquist contour does not encircle the (-1, 0j) point (since in this system there are no poles on the right half plane), as long as there are no poles on the right half plane.

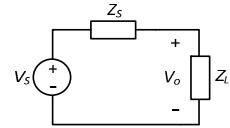

Fig. 1. Connection model of two systems based on their input and output impedances

$$T_{OpenLoop}(s) = \frac{Z_s(s)}{Z_L(s)} \quad (1)$$

When $\|Z_s\| \ll \|Z_L\|$ must be satisfied that

$$|\arg(Z_S) - \arg(Z_L)| \leq 180° \quad (2)$$

For all of the above mentioned it is possible to deduce that in order to obtain stability it is necessary to know accurately the closed loop impedances of all the components of the system.

There are different methods to obtain the closed loop input or output impedance of a power converter such as analytical form or experimental obtaining by means of a Frequency Response Analyzer (FRA) [6], [7]. In recent years other ways of obtaining impedance online have been studied, such as using Pseudo Random Binary Sequence (PRBS) [8], [9].

This article focuses on the study of the accuracy of the closed-loop input impedance model, using the model that assumes that a controlled converter behaves as a constant power load.

This article is organized as follows. Section II presents the complexity of obtaining impedance analysis compared to obtaining the reduced order model. Section III shows the accuracy range of the model for different powers, controllers and components by simulation. Section IV shows the validation of the model through experimental results for real high-power converters.

## II. OBTAINING THE INPUT IMPEDANCE

This section will show how difficult it is to obtain the analytical closed loop input impedance, then present the reduced model to study and present its advantages. In this case the power converter to be studied is a DC-AC three-phase 2L-VSC. To obtain the theoretical impedance is considered a specific case in which the converter has only one current control loop and uses the DQ reference axes.

### A. Analytical Input Impedance

The impedance of analytical form is obtained from the model of the Fig. 2 as it can be seen the total small-signal input impedance ($Z_{iT}$) can be obtained as the parallel of input capacitor impedance ($Z_{Ci}$) and is the closed-loop input impedance of the three-phase VSI ($Z_i$). As shown in (3)

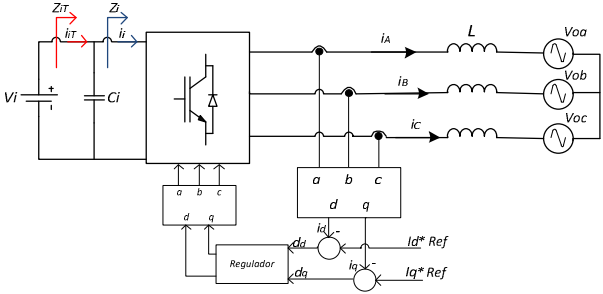

Fig. 2. Three-phase VSI current controlled

$$Z_{iT}(s) = \frac{Z_{Ci}(s) \cdot Z_i(s)}{Z_{Ci}(s) + Z_i(s)} \quad (3)$$

The analytical derivation of the input impedance $Z_i$ can be done using the small signal equivalent model shown in Fig. 3.

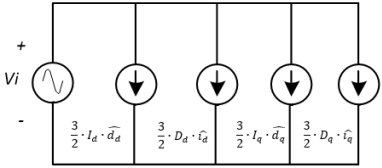

Fig. 3. Small signal input port of three-phase VSI

As can be seen in [10], the control variables DQ can be decoupled, obtaining two first-order systems with a single input and a single output (SISO system). As a consequence, the input impedance of the inverter controlled in DQ in closed loop is determined by

$$Z_i(s) = \frac{2}{3} \cdot \frac{1}{I_d \cdot G_{ddvi}(s) + D_q \cdot G_{idvi}(s) + I_q \cdot G_{dqvi}(s) + D_q \cdot G_{iqvi}(s)} \quad (4)$$

Where $I_d, I_q, D_d$ and $D_q$ are the current and duty cycle magnitudes in the operation point. The transfer functions are defined by

$$G_{idvi}(s) = \frac{1}{sL+r}\left(\omega L G_{iqvi}(s) + V_i G_{ddvi}(s) + D_d\right) \quad (5)$$

$$G_{iqvi}(s) = \frac{1}{sL+r}\left(-\omega L G_{idvi}(s) + V_i G_{dqvi}(s) + D_q\right) \quad (6)$$

$$G_{ddvi}(s) = -\frac{1}{V_i}\left(D_d + G_{PI}(s) \cdot G_{idvi}(s) + \omega L G_{iqvi}(s)\right) \quad (7)$$

$$G_{dqvi}(s) = -\frac{1}{V_i}\left(D_q + G_{PI}(s) \cdot G_{iqvi}(s) - \omega L G_{idvi}(s)\right) \quad (8)$$

As can be seen the calculation of the expression of the input impedance of the three-phase inverter can be quite time-consuming due to the need to solve the system of equations shown above (5-8).

### B. Experimental Frequency Response Analizer

It is possible to obtain experimentally the value of the closed-loop input impedance by means of a FRA, [6]. This method requires very expensive equipment, it is also problematic to measure high voltage and power equipment, because these converters normally have a very large input capacitor, so in order to see the high frequency response is necessary to introduce a disturbance of a very large amplitude in the input voltage.

### C. Analytical input impedance

The model shown below reduces the complexity of obtaining the input impedance. Considering that the inverter behaves as a constant power load (CPL) due to the action of the control loop below its bandwidth, this behavior can be modeled as a negative resistance $R_{CPL}$ which value is given by (9). At high frequencies the behavior is dominated by the input capacitor $C_i$, therefore the total input impedance is given by (10).

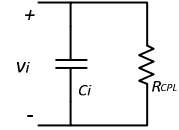

Fig. 4. Proposed model of input impedance of three phase VSI

$$R_{CPL} = \frac{-V_i^2 \cdot \eta}{P_o} \quad (9)$$

$$Z_{iT}(s) = \frac{R_{CPL}}{1 + R_{CPL} \cdot C_i \cdot s} \quad (10)$$

As can be seen, this reduced model has certain advantages over the traditional analytical model, such as the quick obtaining from easily measurable parameters or even those present in the datasheets, avoiding complex analytical obtaining techniques and high-cost measurements, which is especially noticeable in high-power converters.

This input impedance model has already been validated for DC-DC converters [11], [12], and now it is wanted to check its accuracy for DC-AC converters.

## III. STUDY OF THE MODEL ACCURACY

To check the accuracy range of the model several tests with different characteristics have been carried out (using the system present in Fig. 2.) to see if the model fits properly with the real impedance. For all cases, the following considerations have been taken into account when selecting the components:

- The input capacitor has been calculated to have an input voltage ripple of less than 5% of the total voltage.

- For the mains connection filter only one inductance per phase is used, which has been calculated to have a 10% ripple.

- Switching frequency stays constant at 10 kHz

The above considerations have been chosen because they give the smallest possible capacitor in a commercial converter with a correct design. When this capacitor is oversized it adequately filters out the effect of the control on the input impedance. Doing the following studies with the smallest possible capacitor will provide the most critical working point for the model

### A. Comparison of different nominal powers

Below is shown a comparison of the input impedance for 3 different inverters with current control (PI) in the DQ frame with their components calculated for different rated powers. The Fig. 5-7 show both the total impedance ($Z_{iT}$) (parallel of the inverter with the capacitor) and the impedance of the inverter without the capacitor ($Z_i$).
The reduced model has also been added to compare it with the results.

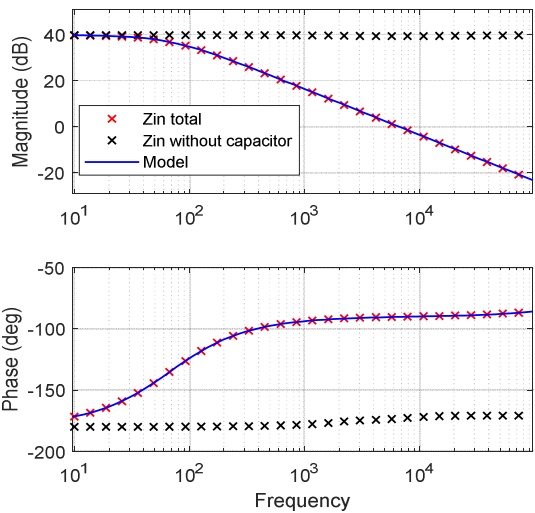

Fig. 5. Closed loop input impedance for 5 kW of rated power.
(*Lfilter=1 mH, Cin=24 µF, PI$_{Gain}$=1, PI$_{TimeCte}$=14.3e-3, R$_{CPL}$=-98 Ω*)

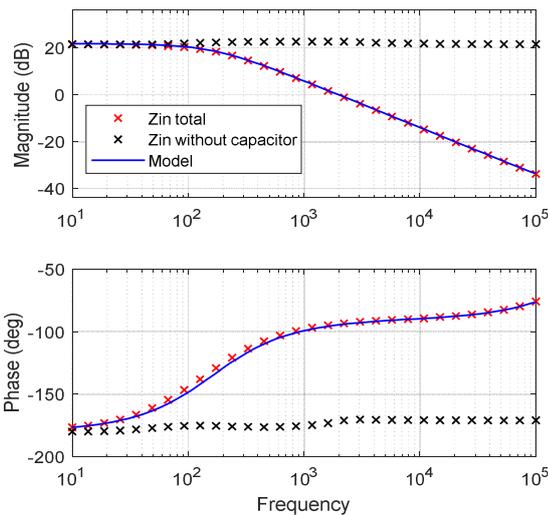

Fig. 6. Closed loop input impedance for 40 kW of rated power.
(*Lfilter=0.3 mH, Cin=80 µF, PI$_{Gain}$=0.3, PI$_{TimeCte}$=4.3e-3, R$_{CPL}$=-12.25 Ω*)

As it can be seen, the model fits perfectly to the impedance obtained by simulation, since the feedforward applied with the input voltage makes the impedance without the capacitor practically constant. So, the model would still adapt if the input capacitor was smaller.

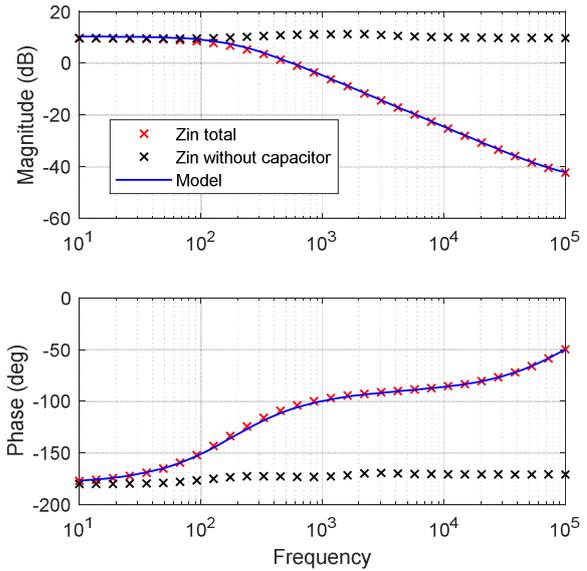

Fig. 7. Closed loop input impedance for 150 kW of rated power.
(*Lfilter=0.06 mH, Cin=270 µF, PIGain=0.06, PITimeCte=8.75e-4, R$_{CPL}$=-3.27 Ω*)

### B. Comparison of different controller bandwidth

For the next study the control bandwidth has been varied to see if it influences the shape of the input impedance. In all cases the converter with a nominal power of 5 kW (Fig. 5) has been used. As can be seen in Fig. 8 when a control with a reasonable bandwidth is used, the model still fits properly to the impedance obtained. On the other hand, when the bandwidth is considerably reduced Fig. 9 it can be observed how the model begins to distance from the obtained result. Although the difference in magnitude and phase is not so significant as to produce instability in the distribution bus due to this small inaccuracy in the model.

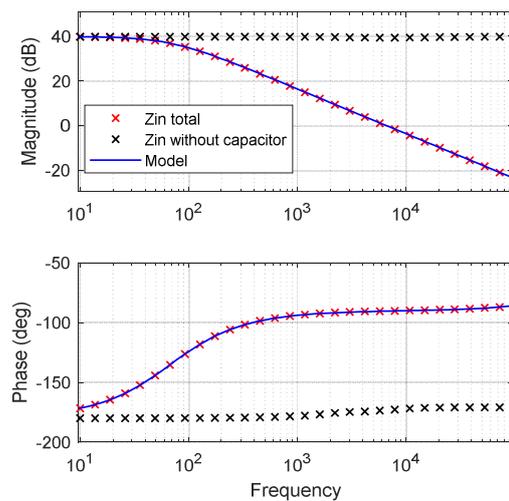

Fig. 8. Closed loop input impedance for 5 kW of rated power and a bandwidth of 160 Hz.

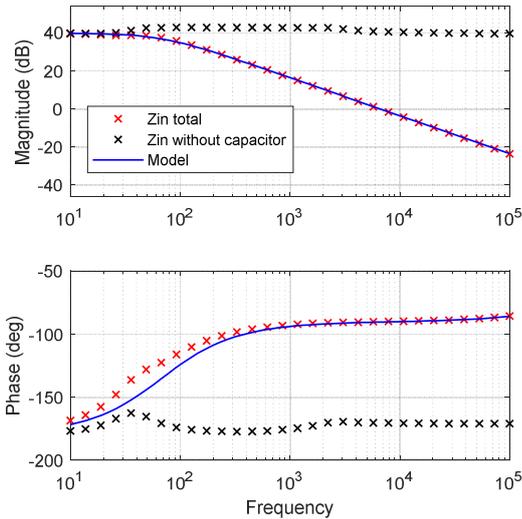

Fig. 9. Closed loop input impedance for 5 kW of rated power and a bandwidth of 15 Hz

## C. Comparison with different reference frame, αβ

In this case, the behavior of the converter is tested when different reference axes are used (αβ). In this case a Proportional-Resonant control with a pair of complex poles conjugated at the fundamental frequency of the network [13] has been used to perform the control.

As can be seen in Fig.10 the model fits with sufficient precision to the results obtained.

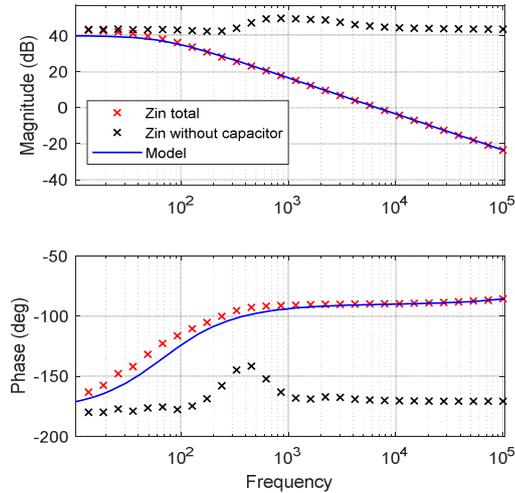

Fig. 10. Closed loop input impedance for 5 kW of rated power using the αβ reference frame

## D. Effect of input voltage feedforward

It has been observed that all the changes made in the previous studies are practically inappreciable with respect to the input impedance thanks to the decoupling of the feedforward from the sensed input voltage. In this study, the feedforward will be modified to see how it affects in the impedance.

As can be seen in Fig. 11 and Fig. 12 the input impedance differs a lot from the proposed model, being more evident when using the αβ reference axes. It is possible to observe how at very low frequencies the constant power load behavior is still maintained, but the fact of making the feedforward constant means that there is no almost constant impedance after the capacitor at medium frequencies and therefore its effect on the total impedance is seen.

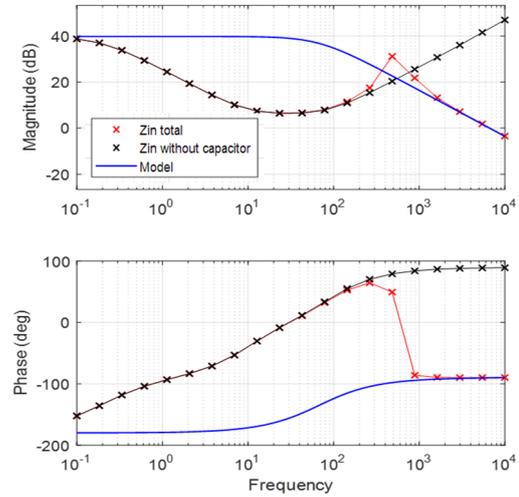

Fig. 11. Closed loop input impedance for 5 kW of rated power. In DQ using a constant feedforward (*Lfilter=1 mH, Cin=24 µF, $PI_{Gain}$=1, $PI_{TimeCte}$=14.3e-3*)

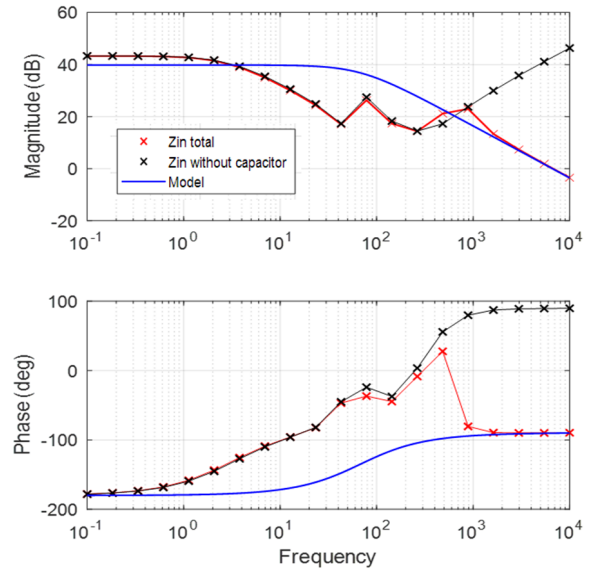

Fig. 12. Closed loop input impedance for 5 kW of rated power. In αβ using a constant feedforward (*Lfilter=1 mH, Cin=24 µF*)

To make a study with a more real case, it has been assumed that in a real design of an inverter has chosen an input voltage sensor without a great bandwidth and is applied a low pass filter in the control to filter noise, finally obtaining a real measure with a bandwidth of 1 kHz.

In the first case, the sensed value of the feedforward has been replaced by a constant value. This study allows us to understand how feedforward affects the system. Later it will be studied what effect the feedforward has on the impedance limiting the bandwidth of the sensed input voltage, something that happens in real converters.

As can be seen in Fig.13 and Fig. 14 the fact of reducing the feedforward sensing bandwidth visibly affects the input impedance, making it different from the model, which may cause instability in the system.

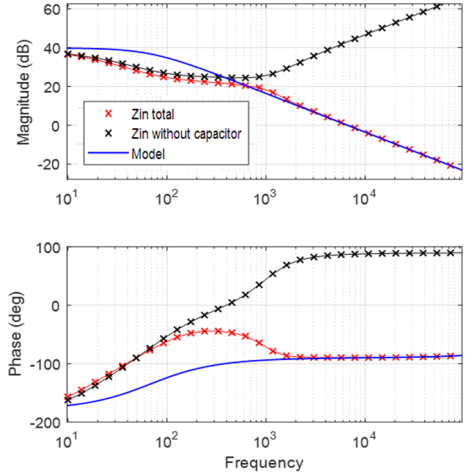

Fig. 13. Closed loop input impedance for 5 kW of rated power. In DQ using feedforward with a 1 kHz of bandwidth. *(Lfilter=1 mH, Cin=24 µF)*

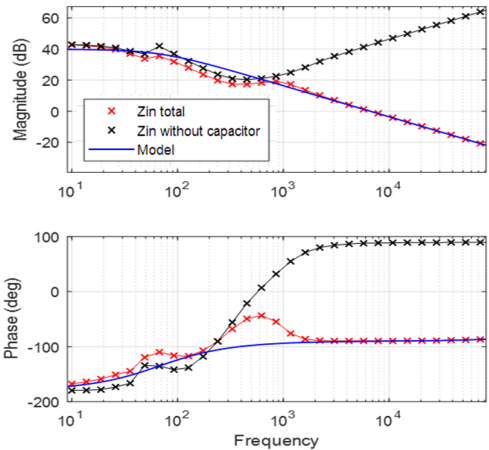

Fig. 14. Closed loop input impedance for 5 kW of rated power. In αβ using a feedforward with a 1 kHz of bandwidth. *(Lfilter=1 mH, Cin=24 µF)*

## IV. EXPERIMENTAL RESULTS

In this section, impedance measurements have been made with a real high power 2L-VSC, and closed loop input impedances have been measured with a method equivalent to that used by commercial FRAs. The simplified connection diagram is shown in Fig. 15 to obtain the impedance a disturbance has been introduced in the DC reference voltage of the EGSTON grid emulator. The DC input voltage in the converter has been sensed with a differential probe, very close to the input of the converter so that it does not affect the inductance of the transmission cables The current from the DC port to the inverter has been sensed with an amperimetric clamp. The results of the measurements have been recorded with an oscilloscope, because it is the instrument that offers more precision and later have been processed by FFT techniques, to obtain the magnitude and phase of the input impedance

Since this is a commercial converter that can handle a power of up to 60 kW, it is not possible to access the current after the capacitor, or modify the input capacitor, which is a little oversized. Unlike the previous studies this converter has an LCL filter at the output, and therefore has implemented in the control an Active Damping. Table I shows some of the parameters of the inverter.

TABLE. I. REAL CONVERTER PARAMETERS

| Parameter | Value |
|---|---|
| Cin | 14.1 mF |
| Rcin | 0.005 Ω |
| Lfiltconv | 0.5 mH |
| Cfilt | 50 µF |
| LfiltGrid | 0.2 mH |

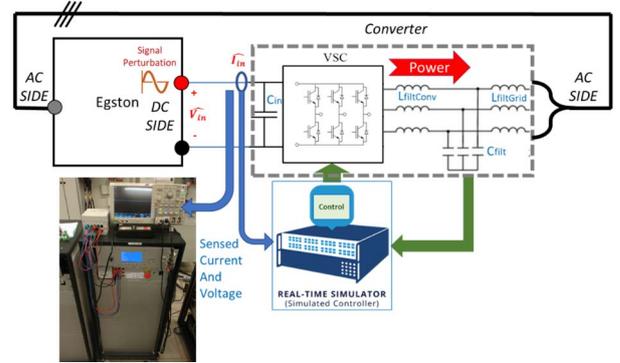

Fig. 15. Setup to perform impedance measurements with the real 2L-VSC.

### A. Comparison of different output powers

In this case the tests were done with a current control on the DQ reference axis, these tests were done for 2 different powers, 2.1 kW and 21 kW. The result can be seen in the Fig. 16.

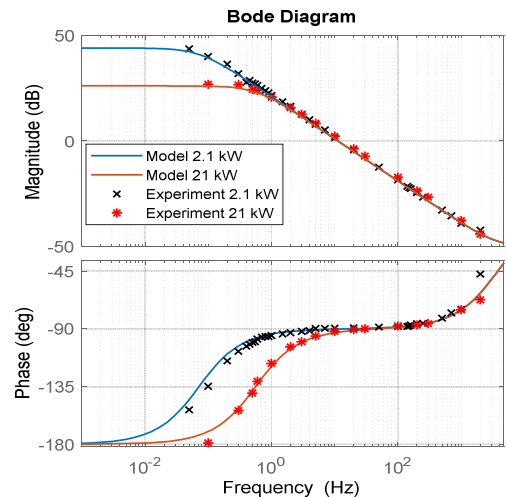

Fig. 16. Experimental results of the input impedance for the same converter working at different powers.

## B. Comparison of different controls

In the following test different controls have been tested, in two of them the time constant τi of the PI controller has been varied. The results are shown in Fig. 17. As it can be seen, the model fits properly to the results obtained for the two PI controls. From the experiments finally shown it can be concluded that if the input capacitor of the converter is oversized the results of the real impedance will match the model.

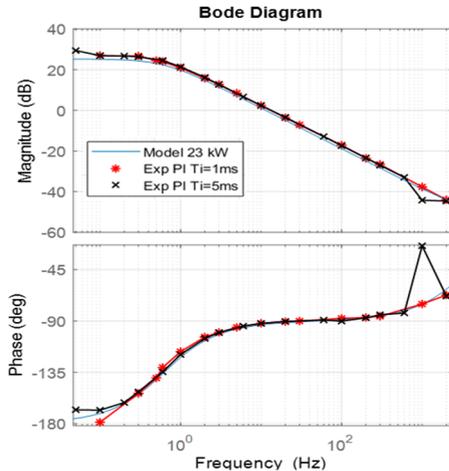

Fig. 17. Experimental results of the input impedance for the same converter working at 23 kW with different controllers.

## V. CONCLUSIONS

In this article the closed-loop input impedance of a DC-AC 2L-VSC converter has been studied for different operation cases, the most critical cases (smaller input capacitor) have been studied by means of simulation. It has been possible to observe how the main factor that influences in the impedance is the feedforward, checking how with a standard bandwidth in the feedforward the input impedance starts to differ quite a lot from the model.

From the experimental results it can be seen how the fact of having an oversized capacitor benefits the precision of the model.

## ACKNOWLEDGMENT

This work is partially supported by the European Regional Development Fund, the Ministry of Science, Innovation and Universities and the State Research Agency through the research project "Modelling and control strategies for the stabilization of the interconnection of power electronic converters" (FEDER/Ministerio de Ciencia, Innovación y Universidades – Agencia Estatal de Investigación/ _Proyecto CONEXPOT-2 (DPI2017-84572-C2-2-R)). It also has been partially supported by the Research Agreement between the Universidad Carlos III de Madrid and the Fundación IMDEA Energía. In addition, it has been also supported by the research project PERSEID (04.022-2018) using the ERIGrid Research Infrastructure and is part of a project that has received funding from the European Union's Horizon 2020 Research and Innovation Programme under the Grant Agreement No. 654113. The support of the European Research Infrastructure ERIGrid and its partner SINTEF is very much appreciated.